\documentstyle[aps,aipbook,floats]{revtex}
\newcommand{\be}{\begin{equation}}
\newcommand{\ee}{\end{equation}}
\begin{document}
\title{The time evolution of GRB spectra by a precessing lighthouse Gamma Jet}

\author{Daniele Fargion$^*$ , Andrea Salis}
\address{Physics Dept. Rome University "La Sapienza"\\
P.le A.Moro 00187\\ 
$^*$INFN sez. Roma1 Univ."La Sapienza", Rome}

\maketitle

\begin{abstract}
Inverse Compton Scattering (ICS) by a relativistic electron beam jet at GeV 
energies (emitted by a compact object as a NS, BH,...), a NSJ, onto thermal BBR 
photons (from a nearby stellar companion) may originate a collinear gamma jet 
(GJ). Due to the binary system interaction the GJ precession would blaze 
suddenly toward the observer leading to a GRB event. The internal GJ cone 
structure is ruled by relativistic kinematics into a concentric onion-like 
sequence of photon rings, the softer in the external boundaries, the harder in 
the inner cone. The pointing and the crossing of such different GJ photon 
rings to the detector lead to a GRB hardness spectra evolution nearly 
corresponding to most observed ones. Moreover expected time integral spectra 
are also comparable with known GRB spectra. The total energy input of tens of 
thousands of such NSJ in an extended galactic halo, mainly cosmic rays 
electrons, should be reflected into the recent observational evidence 
(COMPTEL) of a diffused relic extended halo. Evidences of such precessing jets 
are offered by the discover of galactic superluminal sources, recent HH jets, 
SN1987A outer rings, Hourglass Nebula, planetary Egg Nebula, GROJ1744-28 
binary X-rays pulsar.
\end{abstract}

\section*{}

As in the previous paper we were forced to build up our GJ model by the ICS 
process made by relativistic cosmic rays electron jet onto BBR photons. To 
obtain 
an analytical "direct" formula we first studied the Compton Scattering of the 
anisotropic boosted "BBR" in the electron rest frame and then we transformed 
back in the laboratory frame the resulting ICS output. This approach has been 
successfully tested with the experimental ICS spectrum onto room thermal 
BBR photons performed at LEP [1-2]. However in our binary model the NSJ moves 
in presence of an inhomogeneous and anisotropic flux of photons coming from a 
nearby companion BBR star source. For sake of simplicity (without any loss of 
generality) we first assumed a ring-like photon source within a limited 
incident angle ($\theta_o\sim 45^o$) for the incoming photons. The luminous 
ring plays the role of the companion star, its radial distance from the 
jet, and integral 
intensity have the same behaviour and luminosity of a binary star source at 
the same radial distance. In the ultrarelativistic-Thomson limit the ICS  
differential photon number distribution is [3]
$$
\frac{dN_1}{dt_1 d\epsilon_1 d\Omega_1}=\frac{2\pi\kappa_B T r_o^2 c}{c^3 
h^3} N_o \epsilon_1\int_{\gamma_{min}}^{\gamma_{max}}\frac{\gamma^{-\alpha-
2}}{\beta}\cdot
$$
\be
\cdot\ln\Bigg[\frac{1-\exp\Big(\frac{-\epsilon_1(1-
\beta\cos\theta_1)}{\kappa_B T(1-\beta\cos\theta_{o,min})}\Big)}{1-\exp\Big(
\frac{-\epsilon_1(1-\beta\cos\theta_1)}{\kappa_B T(1-\beta\cos\theta_{o,max})
}\Big)}\Bigg]\cdot\Bigg[1+\Big(\frac{\cos\theta_1-\beta}{1-
\beta\cos\theta_1}\Big)^2\Bigg] d\gamma
\ee
where $\epsilon_1$ is the observed final photon energy, $N_o$ is a normalizing 
factor (consistent with the total flux intensity of GRBs 
$\dot{E}_{\gamma}\sim5\cdot 10^{41}~erg~s^{-1}$ described in eq.4 of 
ref.[5]), $\theta_{o,min}$ and $\theta_{o,max}$ are the nominal 
incident angles between the jet beam direction and the thermal photons. In 
this formula $\theta_1$ (the final external cone angle from the core) evolves 
with 
the GJ sweeping. Indeed the GJ hitting and crossing by the observer reflects 
in the simplest configuration (at small angle approximation) into an angle 
evolution $\theta_1(t)=
\sqrt{\theta_{1,min}^2+(\omega_b t)^2}$ where t=0 corresponds to the maximum 
of the GRB rate and $\theta_{1,min}$ is the observed 
minimal angle from the beam jet center toward the observer. Because of the 
sharp rate decrease at large angle in eq.1 (and eq.3), the approximation holds 
also at large angles $\gamma\theta_1(t)\gg 1$. Overimposed to 
this simplest angular variation one should expect a 
(nearly) periodic "trembling" at millisecond, the inprint of the pulsar jet at 
the characteristic $\omega_{PSR}$ frequency. Moreover an additional nutation 
of the spinning star is possible ($\omega_N\sim\omega_{PSR}\frac{I_\perp-
I_\parallel}{I_\parallel}$) driving the final angle beam $\theta_1$ to a 
fascinating dance along a multiple cycloidal (or epicycloidal) trajectory 
described by the angle $\theta_1 (t)$ written as follows:
$$
\theta_1(t)=\sqrt{[\theta_{1,min}+\theta_{PSR}\cos(\omega_{PSR}t
+\varphi_{PSR})+\theta_N\cos(\omega_N t+\varphi_N)]^2+}
$$
\be
\overline{+[\omega_B t+\theta_{PSR}\sin(\omega_{PSR}+\varphi_{PSR})+\theta_N
\sin(\omega_N t+\varphi_N)]^2}
\ee
where $\omega_b$, $\omega_{PSR}$, $\omega_N$ are respectively the binary 
system, the characteristic pulsar, the nutation NSJ frequencies. The free 
constant 
parameters $\theta_{1,min}$, $\theta_{PSR}$, $\theta_N$ are the minimal 
"impact" angle for $\theta_1$, the maximal angular amplitude of the NSJ 
"trembling" and the corresponding maximal oscillation during the angular 
nutation; $\varphi_{PSR}$ and $\varphi_N$ are the constant phases defined by 
the characteristic initial conditions. The temporal spectral evolution of GRBs 
is given in detail by 
substituting the previous $\theta_1(t)$ behaviour in eq.1. The consequent 
adimensional total photon flux number time evolution, for a monochromatic 
electron jet spectrum, is (after integrating over $\epsilon_1$ in eq.1)
\be
\frac{(dN_1/dt_1 d\Omega_1)_{\theta_1(t)}}{(dN_1/dt_1 d\Omega_1)_{\theta_1=0}}=
\frac{1+\gamma^4\theta_1^4 (t)}{[1+\gamma^2\theta_1^2 (t)]^4}
\ee
The presence of 8 parameters $\theta_{1,min}$, $\theta_{PSR}$, $\theta_N$, 
$\varphi_{PSR}$, $\varphi_N$, $\omega_b$, $\omega_{PSR}$, $\omega_N$ allows to 
fit the many different GRB faces and behaviours [6] as well as to understand 
the 
limiting case: if $\theta_{1,min}$ is large (and so more easely seen the jet 
because of the larger angle area) the average GRB energy is softer (SGR 
event) and less intense. The SGR repetition arises because, in eq.2, $\omega_b 
t$ (for GRB) is the limit function of a more general periodic one (for SGR): 
$\omega_b t\leftrightarrow\sin(\omega_b t)$. Moreover the peculiar case of a 
gamma ray pulsar is recovered from the same expressions when all the binary 
and nutation variability vanish ($\theta_N=0$, $\omega_b=0$) and only the 
pulsar trembling behaviour survives. If, for istance, we parametrize the 
$\theta_1$ angle aperture in eq.2 in Lorentz factor unities, as $\theta_1=
\sqrt{\theta_{1,min}^2+(n/\gamma)^2}$, then the GJ sweeping induces a violent 
"thermal" evolution soft-hard-soft (from eq.1) as shown in fig.1 where tens 
KeV and hundred KeV "thermal" spectra arise for angle evolution at n=1,2,3,10. 

\begin{figure} 
\vspace{2.75in}
\includegraphics{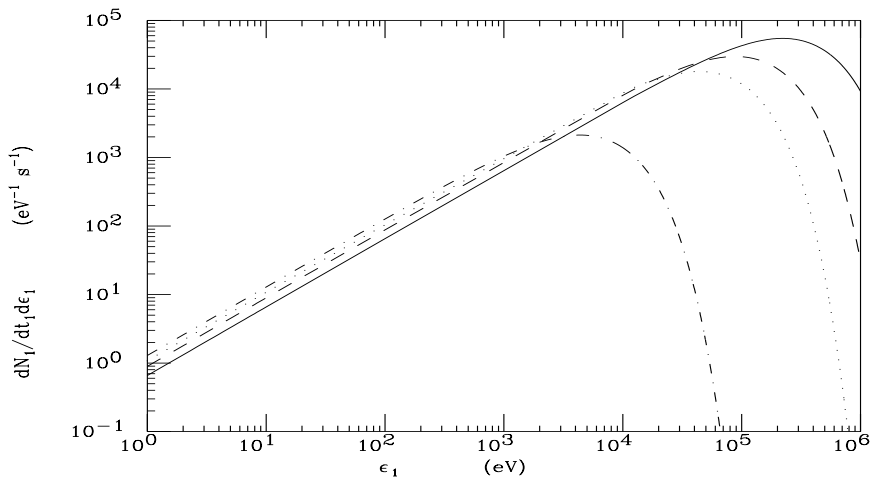}
\caption{ICS energy spectrum, in arbitrary units, for $\theta_1=n/\gamma$ and 
n=1,2,3,10 angle apertures (from continuous to dot dashed curve)}
\end{figure}

For most general $\theta_1(t)$ evolution 
(eq.2) the spectral behaviour is either rapid and trembling (within each GRB 
spike due to $\omega_{PSR}$) as observed and also contains slow thermal 
evolution and repetition within lower epicycloidal frequencies ($\omega_b$, 
$\omega_N$). The final time integral for different $\theta_{1,min}$ angle in 
unidimensional "slices" of these "onion cones" number rate distribution in the 
usual form $F_\nu\nu=\epsilon_1^2\frac{dN_1}{dt_1 d\epsilon_1}$ (from eq.1) are 
shown in fig.2.

\begin{figure} 
\vspace*{2.75in}
\includegraphics{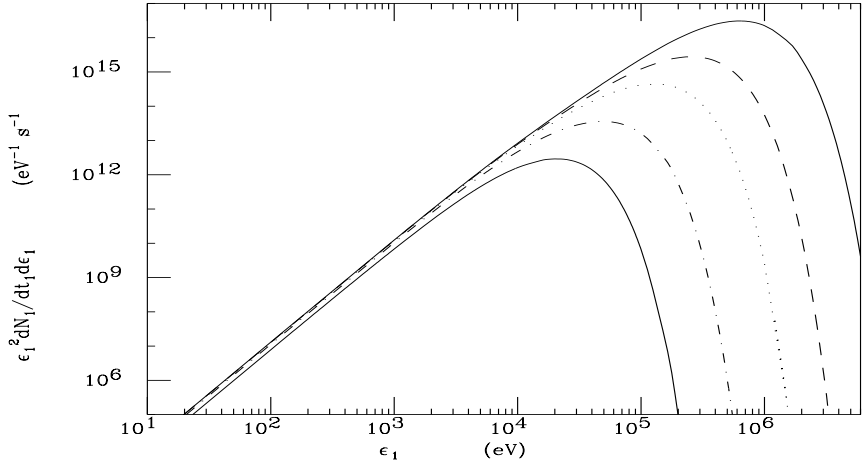}
\caption{ICS integral energy flux, in arbitrary units, for $\theta_{1,min}=
n/\gamma$ and n=1,2,3,5,8 angle apertures (from upper continuous to lower 
continuous curve)} 
\end{figure}

The final integral over a power law electron spectrum fits in a successfully 
way the GRB experimental spectra. For a chaotic "random walk" around 
$\theta_{1,min}$ one may easely approximate the final spectrum by a total 
surface angular integral over $\theta_1$ (see ref.[3] and the successfull fit 
of GRB910601$\_$69736). 
The $\gamma-\gamma$ electron pair production opacity near the jet for the 
parameters given in ref.[3] is within unity and cannot lead 
to thermalization. The isotropy "obsession" of GRB which seemed to push all 
(or most) the theoreticians toward the popular cosmological belief cannot 
force the present GJ 
model toward those huge distances and energies. Indeed an amplification (by 
nearly 10 order of magnitude) of the GJ power ($\dot{E}_{\gamma}$) by a 
corresponding NSJ $\dot{E}_j$ power do {\it increase by 
the same factor the density of the electrons} in the jet (even disregarding 
the copious $e^+-e^-$ pair production), leading to a total screening and 
opacity (like in fireball models) of the GJ and to a final {\it thermal GRB 
spectrum} (in disagreement with observations). Therefore cosmological GJ 
"versions" are not acceptable GRB sources. The total number $N_T$ of such 
active GJ in an extended galactic halo may be roughly estimated by two main 
arguments: (a) the ratio between the jet lifetime 
(assuming a companion solar mass $M_\odot$ feeding by mass transfer the power 
jet) $\tau\sim\frac{E_{SN}}{\dot{E}_j}\sim 3\cdot 10^7 yr$ and the birth time 
of such system (comparable to SN one) $\Delta\tau_{SN}\sim 30 yr$; (b) the 
ratio between the GRB observational probability and the GJ beam solid angle 
size. These GJ numbers are at first approximation respectively one million and 
30000; a corresponding 
source density quite rare in an extended galactic halo. Since the end of 1993 
we were inspired by the above arguments and by just two 
known galactic jet candidates: SS433 and the Great Annihilator IE1740.7-2942. 
In the last three years the candidates and the evidences for precessing GJ 
have blown up. The superluminal sources GRS1758-258, GRS1915+105 and 
GROJ1655-40 whose GJ nature and SGR association has been promptly noted [3]. 
The jet traces in HH34 filaments as in earlier notes [3] became evident 
by last deep Hubble inspections of their jet cores (HH30, HH1, HH2, HH34, 
HH47). Known SGRs have been identified as runaway $\underline{binaries}$ from 
their SNR birthplace with bright infrared companion (somehow obscured by 
dust). The last discover of binary pulsar GROJ1744-28, discussed in ref.[5], is 
an ideal candidate for such precessing GJ. 
The evidence of high velocity NS (HVNS) may explain the isotropization and 
the diffusion into extended galactic halos of the GRB sources, consistent with 
observed isotropy. The GRB flux count break at lowest fluxes may be 
reflected 
into the coexistence of a bounded (homogeneous $\rho\sim cost$) extended halo 
(or corona $<$100 Kpc) and of an evaporating component ($\rho\sim 1/r^2 >$ 100 
Kpc). Last evidences of the twin rings around SN1987A are also favouring 
the presence of two sided precessing jet spraying onto the red giant relic 
shell. One may foresee [4] the presence of a NS relic moving from the SN core 
toward $\underline{South-East}$ (due to offaxis beaming and "rowing" 
acceleration processes). The final discovery (16 January 96) by Hubble of 
the Egg Nebula CRL2688 is probably the most spectacular and detailed view of 
such precessing GJ in space: the outer gas-dust (nearly homogeneous) spherical 
shells allow to diffuse and clearly put in evidence the symmetric sweeping of 
a (twin) conical shape of the precessing GJ in a "frozen" bright, convincing 
picture. GRBs are just similar sources located in extended galactic halo 
sweeping and trembling once the GJ beam cone points toward the terrestrian 
detectors.

\end{document}